\documentclass[letterpaper,twocolumn,10pt]{article}
\usepackage{usenix2019_v3}

\usepackage{tikz}
\usepackage{amsmath}

\usepackage{filecontents}

\usepackage[normalem]{ulem}

\usepackage{graphicx}
\usepackage{amsmath}
\usepackage{amsfonts}
\usepackage{listings}
\usepackage{algorithm}
\usepackage{morefloats}
\usepackage{endnotes,microtype,xspace,graphicx,fancyvrb,multirow}
\usepackage[noend]{algpseudocode}
\makeatletter
\def\BState{\State\hskip-\ALG@thistlm}
\makeatother

\usepackage{float}

\usepackage{mathtools}

\newcommand{\squeezeup}{\vspace{-4.2mm}}

\begin{document}
%-------------------------------------------------------------------------------

%don't want date printed
% \date{}

\title{\Large \bf Binary Debloating for Security via Demand Driven Loading}
%  \title[BlankIt]{BlankIt: Binary Debloating for Security via Demand Driven Loading}
%for single author (just remove % characters)
 \author{
 {\rm Girish Mururu}\\
 girishmururu@gatech.edu\\
 Georgia Institute of Technology
 % Georgia Institute of Technology
 \and
 {\rm Chris Porter}\\
 cporter35@gatech.edu\\
 Georgia Institute of Technology
 % Second Institution
 % copy the following lines to add more authors
 \and
 {\rm Prithayan Barua}\\
 pbarua3@gatech.edu\\
 Georgia Institute of Technology
 \and
 {\rm Santosh Pande}\\
 santosh.pande@cc.gatech.edu\\
 Georgia Institute of Technology
 } % end author

\maketitle
%-------------------------------------------------------------------------------
% Abstract
\begin{abstract}
Modern software systems heavily use C/C++ based libraries.
Because of the weak memory model of C/C++, libraries may suffer from
vulnerabilities which can expose the applications to potential attacks. For example, a very
large number of return oriented programming gadgets exist in {\it glibc} that allow
stitching together semantically valid but malicious Turing-complete programs.  
In spite of significant advances in attack detection and
mitigation, \textit{full} defense is unrealistic against an ever-growing set of
possibilities for generating such malicious programs.
 
In this work, we create a defense mechanism by debloating
libraries to reduce the {\it dynamic functions linked} so that the possibilities of
constructing malicious programs diminishes significantly. The key idea
is to locate each library call site within an application, and in each
case to load only the set of library functions that will be used at that
call site. This approach of demand-driven loading relies on an
input-aware oracle that predicts a near-exact set of library functions
needed at a given call site during the execution. The predicted
functions are loaded just in time, and the complete call
chain (of function bodies) inside the library  is purged
after returning from the library call back into the application.
%In
% order to effectively implement the scheme, we must solve the problem of
% predicting a set of functions at the forthcoming call site that invokes
%the library.
We present a decision-tree based predictor, which acts as an
oracle, and an optimized runtime system, which works directly
with library binaries like GNU libc and libstdc++. We show 
that on average, the proposed scheme cuts the exposed code surface of libraries
by 97.2\%, reduces ROP gadgets present in linked libraries by 97.9\%,
achieves a prediction accuracy in most cases of at least 97\%, and adds a
small runtime overhead of 18\% on all libraries (16\% for glibc, 2\% for others)
across all benchmarks of SPEC 2006, suggesting this scheme is practical.

\end{abstract}
%-------------------------------------------------------------------------------
\section{Introduction}
Modern software relies heavily on libraries that are often built for
supporting a large amount of functionality. In a given application, however,
only a small of amount of such functionality may get used. For example,
programmers leverage Android libraries, lean on machine learning and AI tools, and build on
top of web frameworks to improve productivity~\cite{ANDK,TensorC,Caffe,Flask,Node.js}. Although these
frameworks can be dauntingly large, it is normal to use only a small
subset of the APIs and in fact a small subset of their functionality.
A recent study \cite{217642} shows that only about 10\% of the shared library
functions in userspace programs that ship with Ubuntu Desktop 16.04 are used.
For performance
reasons, these underlying libraries are inevitably written in C/C++. The
memory model offered
by C/C++ suffers from many weaknesses and leads to a large number of
exploits that expose the applications or frameworks that use the libraries, in turn leaving them
vulnerable. One such library that forms the core of C/C++ libraries and
applications is the GNU version of libc, or  ``glibc'' ~\cite{glibc}.
Glibc also acts as a basic building block of other
libraries and their higher-level APIs in GNU/Linux systems.
Unfortunately, the list of vulnerabilities and exploits in glibc keeps
growing, with 99 known and published vulnerabilities at the time of
this writing~\cite{CVE}, of which nine were published in 2018.
%% Can you update the above numbers 

Over the years, several hardware and software defense mechanisms have
been developed to detect and mitigate the attacks; however, such
mechanisms have limitations. Some of the hardening techniques such as
address-space layout randomization~\cite{aslr} (ASLR) and data
execution prevention~\cite{dep} (DEP) cannot protect against advanced
control-flow hijack techniques like return-oriented
programming~\cite{rop} (ROP) and jump-oriented
programming~\cite{jop} (JOP). To defend against the exploitation of
these advanced vulnerabilities, specific
and general mechanisms have been invented. One such well-known defense
mechanism that has been researched extensively is
control-flow integrity (CFI)~\cite{cfi, practical_cfi, control_flow_locking, kcofi, opaque_cfi, uCFI}, a mechanism that attempts to
restrict the program execution to legal paths only.

While coarse-grained CFI such as bin-CFI (a binary-level CFI technique)~\cite{bincfi} restricts jumps to ROP and JOP gadgets at
addresses other than that of functions, it over-approximates the legal
execution paths possible through calls  and returns. This allows
for malicious program paths and thus overlooks certain attacks. To
reduce the approximation of the allowed set of calls from a call site
and to maintain legal returns, a number of fine-grained CFI techniques
such as $\pi$CFI~\cite{picfi} and MCFI~\cite{mcfi} leverage the
source code of libraries. However, computing a fully-precise CFI is an
intractable problem because of the inherent limitations in pointer
analysis ~\cite{control-jujutsu}, irregular control flow in
the presence of setjmp and longjmp, or an inherent inability to
distinguish between a legal and illegal dynamic control flow base with
regard to the current program input.  $\mu$CFI leverages Intel Processor
Trace and static analysis to enforce its unique target code
property~\cite{uCFI} for indirect control-flow transfers. It differs
from our proposed solution in several ways, including:
it relies on Intel PT (which currently requires kernel support);
it handles application code and not libraries;
it requires source;
and it is limited against certain classes of attacks (as we will show
below in a code example).

Control-Flow Bending~\cite{control_flow_bending} (CFB) showed that even
a fully-precise static CFI cannot defend against many non-control data
attacks. CFB demonstrated that with the availability of a dispatcher
function, commonly available in glibc, a CFI mechanism can be fooled to
successfully carry out an attack. CFI also does not handle non-control
data attacks such as certain types of privilege escalation \cite{branch_corr, non_control_data}.
Unlike CFI, which
allows memory corruption but prevents exploits, another class of defense mechanism that prevents memory corruption is memory safety that requires checks
to verify the correctness of memory operations. While memory safety techniques can
thwart almost all control-flow hijack attacks they incur overheads $\geq$
2x~\cite{Nagarakatte:2010:CCE:1806651.1806657}.
Because CFI in general does not completely defend against all the
control-hijack attacks for which it is built, it should be complemented
with defense mechanisms for specific attacks.
Moreover, as defenses strengthen, new attacks are revealed that
undermine the current best practices of defense mechanisms. Due to
the sophistication of attack construction tools, the number of newly
discovered attacks are growing at a much faster rate, which might lead to a losing battle \cite{attack_rate}. 

To reduce the effects of such unseen attacks, a promising way forward
is program debloating\cite{DBLP:journals/tse/ManadhataW11}, that is,
downsizing the amount of code available for constructing an attack
gadget. One of the established ways of accomplishing this is software
debloating, in which unwanted or dead code is removed from the
application or library.  Current compiler-based link-time optimizations
statically determine a set of functions to be removed from the linked 
library using a combined analysis of unreachable functions, global
constant propagation, and interprocedural evaluation of branch
outcomes using the fixed points of unreachable code and
(conditionally propagated) constants~\cite{Debray:2002:PCC:512529.512542,DBLP:journals/toplas/SutterBB05, DBLP:journals/tecs/SutterPCBB07,Le:2015:RSL:2771783.2771785}. Piece-wise compilation~\cite{217642} 
collects address-taken functions of each module and loads only those
functions that are required by the program. However, many functions are still
stitched to the program, so it still relies on a CFI mechanism for defense.
Such approaches retain a conservative over-approximation of the set of
functions that are linked and needed by the program as a whole.
That is, at different program
points, multiple functions of the library could be reachable/needed, so they
are left linked;
This significantly increases the code surface (of dynamically linked functions).
ROP and JOP gadgets across call sites are still available for
exploitation in the code that is linked to the application.

Software debloating at call sites that use function pointers
faces the same limitation of over-approximation seen in CFI, allowing construction
of such gadgets. Although static program debloating is a right step towards thwarting
unknown possibilities of attacks, it is quite ineffective due to the limitations
of static analyzability of programs, because of which a lot of code still
remains linked. A major limitation of the static debloaters is that the 
attacker can replay
the execution of a statically debloated code over and over under the guidance
of a debugger to systematically construct a Turing-complete attack gadget.
Also, most of the state-of-the-art defense techniques in CFI, memory safety,
and software debloating require {\it source code analysis}, thus leaving out libraries
such as glibc that are notorious for extremely complex control flow which poses
a huge obstacle for inter-procedural analysis of reasonable sophistication (static or dynamic) . 

In response to the limitations mentioned above, we propose BlankIt, a
a defense mechanism based on binaries of the libraries that significantly reduces dynamically linked functions, adds protection against ROP and JOP
attacks, and overcomes the limitations of false negative approximation in CFI and significantly diminishes attackers' ability to carry out replay to devise a new attack gadget. 
BlankIt relies on the following core ideas: 
\begin{itemize}
    \item Only load a set of needed/predicted functions from the
    library at a call site on demand at runtime. The predictor acts as
    an input-based oracle and provides a list of such functions required.
    These functions are loaded in a protected read-only area. The other
    functions in the library are blanked out either by writing zeros
    or the semantic equivalent of a NOP.
    \item The execution of the loaded library functions continues under
    the supervision of probes, which fire before each function's
    invocation; if more functions are needed for execution, they are
    loaded on demand under the guidance of an audit and alarm phase.
    \item When the library execution returns to the call site, the
    complete call chain of loaded functions is purged or blanked out.
    \item On misprediction, a simultaneously running process receives the arguments
    to the library function to run with full memory safety and check for violations.
\end{itemize}
The above scheme significantly diminishes an attacker's ability to
create a gadget that transcends the set of functions loaded at a call
site. For example, if the attacker attempts to jump to a location of 
any ROP or JOP gadget that is inside a function which is not loaded, the
control flow will go to a blanked section (i.e. the instruction pointer
will find only zeros) and crash. Blanking and loading just-in-time
disallows a return instruction to target all possible return addresses,
unlike in bin-CFI~\cite{bincfi}, which makes BlankIt a stronger binary control-hijacking
defense mechanism.
Another binary-level tool called TypeArmor~\cite{typearmor} tries to
handle this, but it essentially appends a feature to
bin-CFI that reduces only the over-approximation of forward edges.
Also, BlankIt's zeroing mechanism does not make any assumptions about the
binary for collecting the legal targets, as in the case of bin-CFI ~\cite{bincfi}.

The success of the blanking scheme depends on predicting the exact set
of functions needed at runtime. Over approximation expands
dynamically linked functions (e.g. leaving more ROP gadgets intact), whereas under
approximation increases runtime overheads. BlankIt prediction is
highly precise, relying on
machine learning to determine the input-related dynamic execution path
in an oracular manner. The learning technique overcomes the
limitations of over-approximation that exist in the case of
fine-grained CFI mechanisms like
$\pi$CFI or MCFI. For example, in the general case like that of Code
\ref{code:foo}, any fine-grained
CFI mechanism  or software-debloating mechanism like piece-wise compilation~\cite{217642} fails to see
that \texttt{myFnPtr} always calls function
\texttt{foo} when \texttt{argc} is greater than 2 and
classifies both \texttt{foo} and \texttt{bar} as valid jumps from the
function pointer call.
In contrast, our prediction mechanism, based on profiling several
call characteristics, learns that \texttt{foo} is called by
\texttt{myFnPtr} when \texttt{argc} is greater than 2. Under
\textit{perfect} prediction or oracular behavior (which we will show is
achievable), BlankIt does
not allow \texttt{bar} to be loaded. If mispredictions may occur and
are allowed, then BlankIt will not load \texttt{bar} without entering a
audit and alarm phase that checks for violations. Similarly, handling the valid set of jumps in the
case of \texttt{setjmp}/\texttt{longjmp} is very easy in BlankIt compared to
CFI mechanisms.

Code pointer Integrity (CPI)~\cite{Kuznetsov:2014:CI:2685048.2685061} is a defense
mechanism that uses memory safety only on pointers that can directly or indirectly modify
code pointers (control-flow) thus lowering the overheads of a full memory safety
mechanism but allowing some control-flow bending attacks. For example, in the non-control
data attack from~\cite{branch_corr} shown in Code \ref{code:attack}, the user is verified
by an input password, and the result is assigned to the variable "user". The user is
frequently checked to provide access to certain operations. However, inbetween the
checks, the function interacts with the user to get  some other inputs. Input "someinput"
can be manipulated by the user, causing a buffer overflow vulnerability in line 12,
and leading to super user access. This attack is not handled by CPI or the latest
CFI ($\pi$CFI, $\mu$CFI), but in the case
of BlankIt, the "call\_super\_user" function is not predicted, and the attack is detected in the
audit and mitigation phase that checks for violations (please refer to section ~\ref{misprediction} for details). However, BlankIt only handles such attacks when a
function call is involved, since that is the prediction granularity it currently operates on. On top of flagging such attacks, another advantage is that
BlankIt 
is a binary defense mechanism and works on glibc, unlike most of the state-of-the-art defense mechanisms like CPI, $\pi$CFI, or CETS. 
\lstset{language=C,
basicstyle=\footnotesize,        % the size of the fonts that are used for the code
breakatwhitespace=false,         % sets if automatic breaks should only happen at whitespace
breaklines=true,                 % sets automatic line breaking
captionpos=b                    % sets the caption-position to bottom
keepspaces=true,
numbers=left,
showtabs=false,                  % show tabs within strings adding particular underscores
tabsize=2,
numbersep=2pt,
}
\squeezeup
\renewcommand{\lstlistingname}{Code}
\begin{lstlisting}[xleftmargin=.1\textwidth,caption={Snippet for considering CFI and BlankIt approaches},label=code:foo]
void foo(int M, int N)
{
	...
    for(int i = 0; i < N; i++)
    {
        A[i] = A[i-1] + A[i-2];
    }
}

void bar(int a, int b)
{
    ...
}
typedef void(*pt2Func)(int, int);
void main(int argc, char **argv)
{
    pt2Func *myFnPtr = &bar;
    int a = 0, b = 0;
    if(argc > 2){
        myFnPtr = &foo;
        a = 1; b = 2;
    }
    myFnPtr(a, b);
}
\end{lstlisting}
\squeezeup
\begin{lstlisting}[xleftmargin=.1\textwidth,caption={Attack missed in CPI and $\pi$CFI but caught in BlankIt},label=code:attack]
void Access(char pwd[20])
{
    char str[SIZE], user[SIZE];
    char *someinput;
    
    verify_user(user, pwd);
    if(strncmp(user,"admin",5)){
        ... 
    }else{
        ...
    }
    strcpy(str,someinput);
    if(strncmp(user,"admin",5)){
        call_super_user();
    }else{
        call_normal_user();
    }
}
\end{lstlisting}

For successful deployment of this technique, the prediction
accuracy must be very high, and for SPEC CPU 2006 benchmarks our
technique achieves an average prediction accuracy of
over 94\% (100\% on 3 benchmarks, and over 97\% for 9 benchmarks).
BlankIt on average reduces the number of dynamically linked functions by over 97\% and
reduces ROP gadgets in the code section by over 97\%. BlankIt incurs a small overhead of 18\% on SPEC 2006 with all libraries.
%% COMMENT: Update the performance number - overhead for all libs, full SPEC 

\subsection{Contributions}
The following are the contributions of this work:
\begin{enumerate}
    \item A framework for dynamically loading and unloading library functions on demand at
     binary level.
    \item A prediction mechanism for predicting the required library functions to be 
    loaded at each call-site, thus avoid over-approximation (such as in CFI).
    \item An audit and alarm technique that on misprediction runs a full-blown memory safety
    check on the library function to catch any attack.
    \item An evaluation of our framework using specCPU2006 benchmarks and
    its supporting libraries--glibc, libm, libgcc, and libstdc++--some of which are
    are not handled in mechanisms that depend on library source.
\end{enumerate}
Our contributions result in the following security implications:
\begin{enumerate}
\item The set of active dynamically linked functions is reduced to protect against future,
unforeseen exploits that mainly rely on code reuse and replay-based attack construction,
and we provide reasonable metrics and results for
the same.
\item The glibc library, notorious for vulnerabilities and not tackled
by most of the state-of-the-art defense techniques, is addressed.
 \item Safety guarantees for \textit{known} exploits that
 state-of-the-art CFI techniques cannot provide (as seen in the function
 pointer example in Code \ref{code:foo}, or the example case study in section ~\ref{case_study}).
 \item Some of the control-flow bending attacks that depend on non-control data
 are thwarted, which falls under the area of attacks deemed as outside the scope of CFI
 and also CPI.
% \item In the worst case, the proposed technique can switch into a
% mitigation phase, which can manifest itself as a program abort, an
% alarm, or sandboxing mechanism (all out of reach for current CFI
% techniques when execution is within an over-approximated function call
% set). There is the orthogonal choice to apply CFI during this phase, as
% well.
% \item Program completeness is still guaranteed via demand-driven
% loading, even though program parts are cut from the image by a machine
% learning-based approach (which is inherently uncertain). Thus, the
% usability under normal execution conditions (misprediction but no
% attack) is guaranteed.
\end{enumerate}

\section{ BlankIt Overall Framework   }
BlankIt is an environment for demand driven loading of library functions at runtime.
BlankIt first creates a dynamic environment for demand-driven loading when
the library is initially imported. Then during runtime it loads
library functions on demand. When the library is imported, BlankIt does the
following for every function:
inserts a probe at the function entry,
copies the function code after the probe into a safe read-only memory,
and wipes (blanks) out all the code after the inserted probe.

% \begin{figure}[h]
%     \centering
%     \includegraphics[scale=0.3]{fig/blankit_app_and_cachemanager.png}
%     \caption{A high-level view of the runtime system%. An application interacts with a manager that loads and purges library functions on a call-site-by-call-site basis.
%     }
%     \label{figure:blankit_app_and_cachemanager}
% \end{figure}

%The runtime, shown in Figure
%\ref{figure:blankit_app_and_cachemanager}, represents the dynamic
%library function loader as a cache manager that copies back (unblanks) the
%function from the library cache into the library on a prefetch
%(predict) directive. %And as indicated in Figure
%~\ref{figure:blankit_link_delink},
The application itself is also instrumented beforehand. A predictor (a decision tree
learnt through off-line profiling), is instrumented at various points in the
program's control flow graph prior to each library call site. %It
% is hoisted in order to place it where the inputs to the
% predictor are first available.
Its output at runtime is simply the chain of functions within the library that it expects to occur
at a given call site.
Different artifacts are included in the
decision tree, such as the call site ID, reverse dominance frontier (RDF),
and the arguments to the library function. 
When the application calls a library function, the probe (inserted
when the library was imported) gets invoked. This probe checks if the functions
called are as predicted, and if they are, execution continues inside loaded functions.
If, however, the call is not through legitimate means (non-predicted
and not legal) then an attack is detected. Any attempts to bypass the
probe results in a fault or crash because the code section has been blanked.
Finally, when the control returns to the application, the functions that
were copied are reset back to zero.% as shown by the ``blank'' call in Figure \ref{figure:blankit_link_delink}.
This purging is what minimizes dynamically linked functions.

% \begin{figure}[h]
%     \centering
%     \includegraphics[scale=0.3]{fig/blankit_link_delink.png}
%     \caption{The flow of execution through a program. The flow is unknown before running (red, upper right) but has some concrete path at runtime (orange, bottom right). BlankIt handles this by loading libraries into a cache at start-up and essentially linking and de-linking them just before and after each function use (green path, left).}
%     \label{figure:blankit_link_delink}
% \end{figure}

\begin{figure}[h]
    \centering
    \includegraphics[scale=0.4]{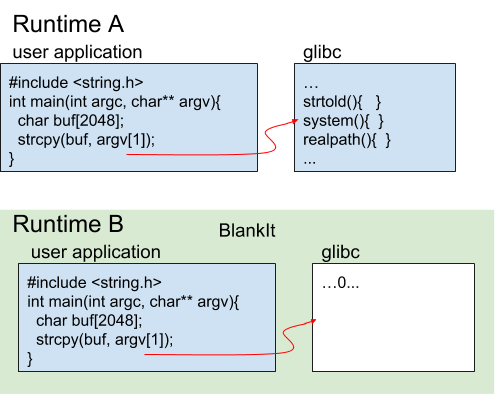}
    \caption{An application running with (B) and without (A) BlankIt (showing a basic return-to-libc attack from \cite{expressiveness_retlibc}).}
    \label{figure:blankit_mot}
\squeezeup    
\end{figure}

The defense of BlankIt at a higher level is depicted in Figure \ref{figure:blankit_mot}.
It shows that in the normal runtime A, an overflow vulnerability can be
exploited to jump to an arbitrary address within glibc, but in runtime
B with BlankIt, since the library functions are wiped out, jumping to an
arbitrary address results in a fault. The prediction mechanism
reduces the set of functions that can be reached through the
vulnerability.

As mentioned, a highly accurate yet light-weight prediction framework is very critical to the
success of BlankIt. The prediction framework
first classifies call chains inside the linked
library as divergent or non-divergent. This is based on the underlying
call graphs. As much as 27\% of the functions within glibc are found to
exhibit linear control flow (i.e. all functions are must-reachable);
these can be loaded without performing any
dynamic prediction. The remaining functions (73\%), however, must be
dynamically predicted. The prediction models are built in a
context-sensitive, deliberate manner during training runs. Not all context
is considered. That is, generating a model that captures multiple
features of a library call site's context (such as preceding program
paths, call chains, etc.) could provide high accuracy,
but the elaborate model could also pose very high overheads.
Instead, we compute the predicates on which the call site and call arguments are
control dependent and generate the model solely based on that. More details
on this are presented in the forthcoming sections.

We empirically show (based on the
intuition of static value separation via
SSA) that a model based on reverse dominance frontiers is accurate and
lightweight. Such an approach provides a highly likely subset of the
may-reachable set of functions from a given call site and the subset itself is a function
of arguments of the callsite summarized by the static-dynamic artifacts
in the application. 
All of this together allows us to predict the may-reachable set of
functions  based on the calling context at each call site.

When a function that is not in the predicted set is invoked at runtime, it means either there was a
misprediction or an attack. In the case of misprediction, the function was required but left blank. It can also happen, however, that a function is part of the predicted set but not actually needed. In this
latter case,  the execution continues without any glitches, but it exposes additional function
surface. Overprediction, which is very rare, as shown later in Section \ref{sec:evaluation}, is
still better than static approximation of reachable functions in CFI, as only valid call chains are
loaded. In the case of underprediction, i.e. when functions are required but remain blanked,
BlankIt starts an audit and alarm phase to check for an attack. In this  phase the library
function and its arguments are handed over to a process that runs the library function with a full
memory safety mechanism under Valgrind~\cite{Nethercote:2007:VFH:1250734.1250746}, which then checks
if the misprediction is an attack.
% BlankIt can handle this in multiple ways such as by waiting for Valgrind
% to report if there was an attack or by continuing execution and letting Valgrind to report an 
% attack to a cleaner which can then restore the changes by the application or killing the process on
% an attack. The choice depends on the process and the execution environment. For example, a virtual
% machine environment can easily restore the changes made by an application.   

% thus detecting any exploits.
% For an exploit prevention mechanism rather than detection mechanism, we have to wait for the result of the full memory safety run.  

\subsection{Threat Model}\label{sec:threat_model}
In our threat model, we assume that the program is not self-modifying,
and an attacker can read/write the data section and read/execute the
code section of a vulnerable program. We assume that the application
source, LLVM compiler that instruments prediction (decision trees) in
the application, the dynamic BlankIt runtime, and the
underlying hardware are not compromised. The arguments to the library
calls can be tampered before the call but after the earliest available
points in the application.
%We assume the BlankIt runtime is a
%protected environment and has sandboxed the user application; as such, 
%the prediction sets are tamper-proof.
Prediction calls to BlankIt from user applications are protected against any
argument tampering using pass-by-register semantics; further, these arguments
are guaranteed to fall within a compile-time known prediction set.
Application code pages are W $\oplus$ X, and BlankIt maintains this
property between copying and blanking operations to guard against code
injection (and potentially probe insertion).
All external libraries can be a source of threat, and any such external
call must be protected.
We make no assumptions on obfuscated code or hand-written binaries or
any other calling conventions, unlike some CFI mechanisms.

Repeatedly calling BlankIt's copy and blank
functions does not change the state of the application nor increase dynamically linked functions.
Attackers are unable to hijack the loading of the functions. Potentially there are two possibilities here:
one is to influence decision trees and the second is to take over the probes. Both are impossible.
The decision trees are embedded and compiled into application code (SPEC) and use read-only arguments
imported from the application; since the attacker has no access nor the ability to tamper the application code,
she can not maliciously influence the decision trees.
BlankIt probes cannot be hijacked or triggered by
an attacker either: such an arbitrary probe would be caught the same way any
arbitrary jump is caught (due to illegal control flow). Secondly, probes are inserted at the respective
call sites in the application and are compiled into the same. Thus, the attacker has no  ability to tamper the code at the probe.
%% COMMENT: Guys can you check the above statements about why probe cant be hijacked? 

% Notes from santosh:
% - external components (linked) are a problem. we have a trusted copy of them, and we are using the trusted copy.
% - The compiled components are trusted.
% - The predicted set is tamper-proof. read-only.
% - Can we make any claims about tamper-proof arguments when calling probes?
%   We need to guarantee that the hacker cannot push a different index into the probe call.

% - other things that might help with the writing:
%   - Trusted platform module. TPM
%   - check santosh's micro papers for authentication tips
%   - can try googling "runtime certification of untrusted modules" or things like this
%   - level 0, kernel-related stuff
%   - Detect tampering by a merkle tree. (doesn't stop but can at least detect)
\section{ BlankIt Runtime}\label{sec:BlankIt}
For demand-driven loading, we have developed BlankIt, a binary tool based on
Intel's Pin~\cite{Pin}. It wipes (blanks) out all functions in a library and loads (copies) them only on demand.
The details of the design and optimization of BlankIt are described here.
The runtime is guided by the predictor which is described in the next section. 

\subsection{BlankIt Design}
BlankIt is a pintool and adheres to the programming model set forth by Pin. 
There are two modes in which Pin operates: JIT and Probe. JIT mode is
flexible but slower; probe mode is limited but faster. Probe
mode supports instrumentation at the start and end of functions, but not
at instruction granularity. In probe mode, all the instrumentation is
inserted at the start-up, and then execution is handed off to the
application to run natively; thus, the overhead incurred is the probe
insertion overhead at the beginning and the dynamic execution costs of
executing probes. Since the instrumentation is
static, there are no stalls during the execution, and an attacker cannot
use this to his/her advantage. Due to these reasons, BlankIt uses probe mode.
In terms of functionality, probe mode is sufficient for BlankIt, and in terms
of overhead, it is necessary.

On-demand loading is achieved in BlankIt through two stages. The first is
when a library is loaded initially for execution. The second is during
execution of the application. At initialization time, BlankIt iterates over
all of the executable's shared objects. Then for each function within
a shared object, it overwrites the first few bytes with a trampoline and
sets the remaining bytes to zeros. In other words, once initialization
is complete, every shared object's text
section has been wiped out and replaced with thunks. BlankIt
maintains a separate copy of functions in a trusted cache.
At runtime, the application normally calls into its library functions without any
changes. The original functions now contain trampolines at their start, however, which bounce
execution into a generic handler, or in Pin-speak, a probe. At this point, BlankIt has the
responsibility of patching back the original code and transferring control accordingly. The
BlankIt probe copies the original function's bytes back into place (i.e. from the end of the
trampoline to the end of the function) and then returns to the function. This mechanism is illustrated below.

Figure \ref{figure:probe_copy_diagram} depicts this mechanism for a
two-function call chain. When the application calls a library function
(\texttt{malloc} in this case), execution proceeds along arrow 1 to the
original function location in libc. During initialization, however,
BlankIt had replaced it with a trampoline. The trampoline causes execution to
flow along arrow 2 to the \texttt{probe\_copy} function. Focusing first
on lines 6-9, a loop over a map \texttt{m} copies and remembers all
functions of the currently predicted set of functions. Here we see that
\texttt{malloc} and \texttt{mmap64} are needed, so they are copied back
into place in libc. Arrows 4, 5, 6 and 7 are needed because in this
example, \texttt{malloc} depends on \texttt{mmap64} for this particular
invocation.

\begin{figure}[h]
    \centering
    \includegraphics[angle=0,width=0.50\textwidth]{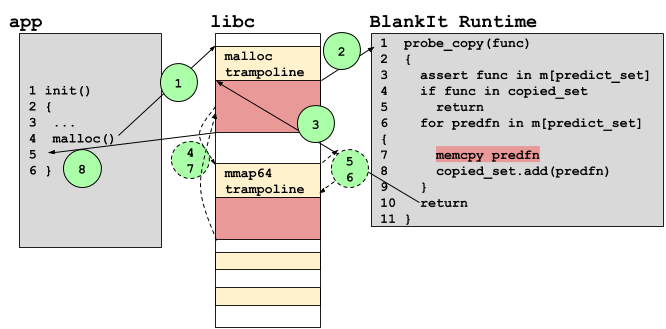}
    \caption{An example of how the copy probe works. Execution flows in order along the arrows.}
    \label{figure:probe_copy_diagram}
\squeezeup
\end{figure}

There are two other important features of Figure \ref{figure:probe_copy_diagram}. The first is that the trampoline in \texttt{mmap64} is still taken (the two small dashed arrows at 5 and 6), even though its code is already in place and no additional copying was needed. This is handled in lines 4-5 in \texttt{probe\_copy}. The second is that there is no blanking at the end, i.e. arrow 8 proceeds back to the application. These two details are treated in the discussion on optimization.

\begin{figure}[h]
    \centering
    \includegraphics[angle=0,width=0.50\textwidth]{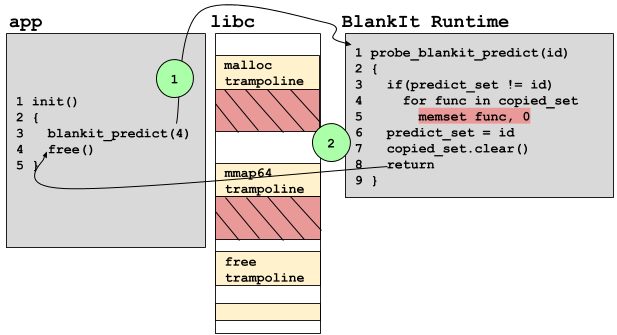}
    \caption{A prediction call that passes an ID of 4 for the upcoming predicted call chain. Within the prediction probe, the last chain of exposed functions will be blanked if the predicted set has changed. In this case, it has changed (from \texttt{malloc} and \texttt{mmap64} to \texttt{free}), so blanking occurs.}
    \label{figure:probe_blankit_predict_diagram}
\squeezeup
\end{figure}

A second type of probe (Figure
\ref{figure:probe_blankit_predict_diagram}) is used for both blanking
and prediction. In a BlankIt-enabled application, prediction calls are
hoisted before the library calls. The predictions are
\textit{themselves} library calls, though, and BlankIt simply replaces
them with a probe during initialization.

In Figure \ref{figure:probe_blankit_predict_diagram}, the application
calls a library function (\texttt{free}), so at compile-time, a call to
\texttt{blankit\_predict} was inserted with an ID of the
profiled/learned call chain at that call site. The initialization for
BlankIt replaced all \texttt{blankit\_predict} calls with
\texttt{probe\_blankit\_predict}, which allows it to blank the last set
of copied functions (lines 4-5) and update the ID of the new prediction
set (line 6). \texttt{probe\_blankit\_predict} then returns along arrow
2 to the application, where it invokes the library function,
\texttt{free}. At this point, the behavior is similar to that described
in Figure \ref{figure:probe_copy_diagram}.

\subsection{BlankIt Optimizations}
A number of optimization opportunities are available in runtime. We
focus on just two of several that were implemented.
%and preloading
%prediction chains.
Some optimizations were ineffective, as well. For example,
prediction allows for the possibility of copying not only the missing
bytes of a function back into place but also overwriting the
trampoline, as well. The upshot of this is that there will be no
trampoline overhead for subsequent calls in the predicted chain (i.e.
arrows 5 and 6 in Figure \ref{figure:probe_copy_diagram}). The
potential downside is that the trampolines must be copied \textit{back}
when the blanking is done. There was no noticeable runtime difference
with this change.

% \squeezeup
\subsubsection{Lazy Blanking}
Figure \ref{figure:probe_blankit_predict_diagram} depicts the second
type of probe that is responsible for storing predictions and blanking
the previous ones. As discussed, blanking does not happen at the return
of every function call. Rather, it happens from within the next
prediction call that occurs just before the first/entry call to a
library. An alternative could be to blank at the return of every
library call, but this has much more call overhead. More importantly,
because blanking occurs when execution re-enters a library, there is a
chance that the predicted functions are the same as the last predicted
set. When this is the case, BlankIt does not need to blank anything. This
kind of ``lazy blanking'' trades some security (by allowing the last
call chain in the library to remain unblanked and exposed until the
next library call) for speed; it still enforces security across call
sites, though, since no two call chains remain linked to the application
at the same time. Thus, this optimization does not have any adverse
effect on attack surface or ROP gadgets as far as the library is
concerned.

% \squeezeup
\subsubsection{Full Call Chain Loading}
BlankIt can be enabled to predict the entire sequence of calls within
a library call from the application. This sequence will facilitate
unblanking the next function and blanking the previous
function in the call chain thus securing all functions in the library
except for the current executing function. This will incapacitate any 
ROP or JOP attack within the call chain. Note that pin allows for calling
a probe after every routine call during which the previous function is
unblanked. However, by not blanking the previous function we expose the
code within the loaded call chain. If an adversary would use any gadget
within the loaded call chain its only a matter till the next jump to a 
function or blanked region is performed when the attack is caught. Any ROP
within the call chain can also be caught by a shadow stack such as Intel's 
CET~\cite{intel-CET}. As is reported later in Section~\ref{sec:evaluation}, the number of
maximum gadgets exposed at any library call advocates this optimization.

\subsection{Audit and Alarm: Handling Misprediction}\label{misprediction}
The drawback of any machine learning mechanism is that the deployment 
is not always 100\% accurate. Such a reduction in accuracy results in
false positives in BlankIt. However, when a mispredict occurs,
BlankIt enters an audit and alarm phase, in which the library function arguments and the
function name are passed on to an auditor process. The auditor loads the library
through a dlopen call and calls the library function while executing with
Valgrind~\cite{Nethercote:2007:VFH:1250734.1250746}, a complete memory safety mechanism.
Further, BlankIt can choose to wait for its auditor to report an attack,
or  continue execution and let the auditor restore changes by the application, or
kill the process.
The choice depends on the process and the execution environment. For example, a virtual
machine environment can easily restore the changes made by an
application~\cite{Zhang:2013:OVC:2535461.2535463}.

% TODO cporter: Usenix 2019 submission. Tell the professor to look at this old version of this
%               paragraph and compare it with the new one. The old one was a weaker claim, because
%               it assumed the arguments could be tampered with. But the threat model assumes the
%               application is safe. Thus, the auditor process should never received corrupted
%               arguments.
BlankIt also handles a few cases of non-control data attacks within the application other than the kind
shown in Code \ref{code:attack}. Please refer to the appendix section~\ref{sec:nc} for more details.

% or if the
% accuracy drops below a threshold, BlankIt can raise an alarm to notify that
% an unexpected behavior occurred. Such alarms are not possible in
% over-approximating CFI mechanisms, whether coarse- or fine-grained.
% Once an alarm is raised, BlankIt enters a mitigation phase, in which BlankIt can
% be complemented with a fine-grained CFI mechanism to load functions
% only when they are at least deemed legal by CFI. Note also that by
% design BlankIt has on-demand loading, and so a mispredict will not
% necessarily crash the program because of some missing, debloated
% component. For practical use this is critical. And if the highest
% possible security is needed and a mispredict ought to cause the program
% to abort,then BlankIt can still entertain this functionality.

\begin{figure}[h]
    \centering
    \includegraphics[angle=0,width=0.50\textwidth]{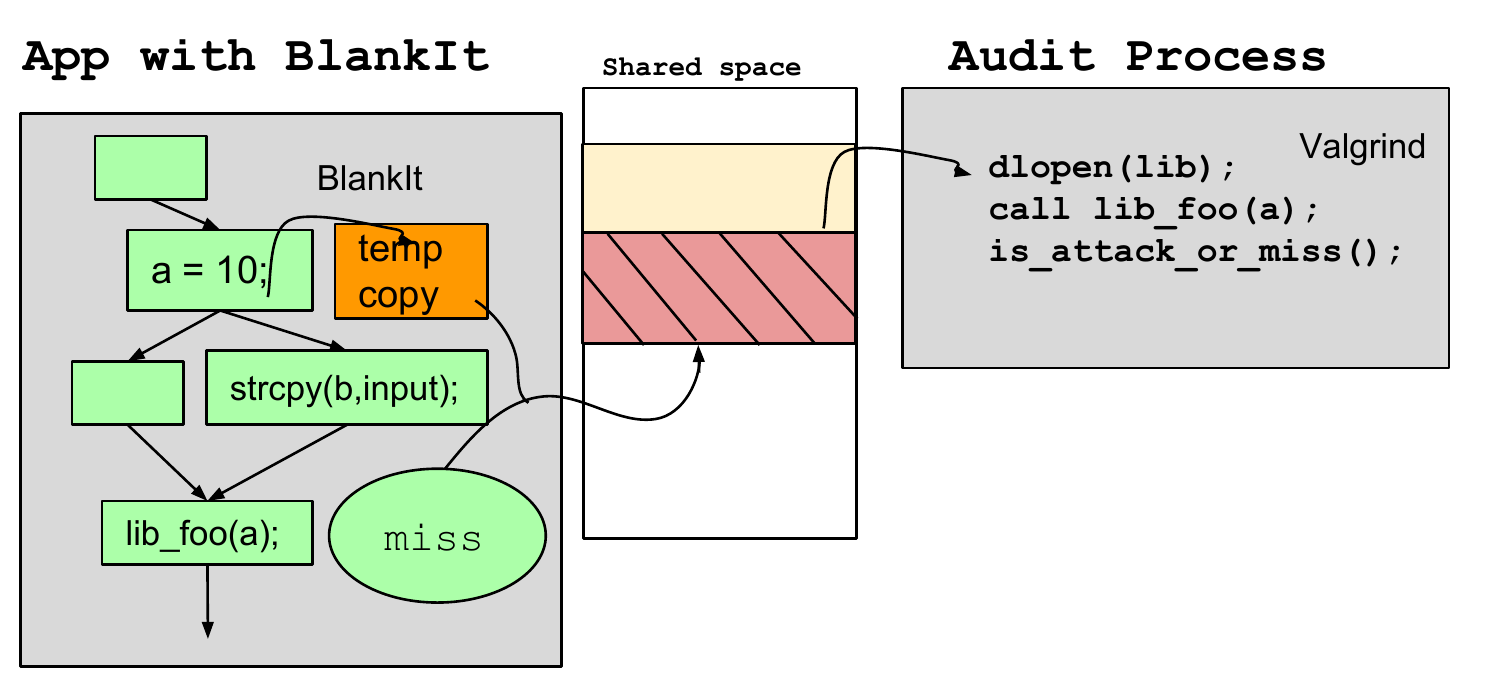}
    \caption{ Audit phase either on misprediction or on an attack. BlankIt handles
              a few cases of non-control data attacks within application.}
    \label{figure:misprediction_phase}
\squeezeup
\end{figure}
\section{Call Graph Prediction}

% \begin{figure}[h!]
%       \centering
%         \includegraphics[scale=0.20]{fig/model_overview.eps}
%         \caption{Overview of the prediction framework}
%         \label{figure:framework}
% \end{figure}

The BlankIt runtime is in some sense driven by the prediction part. The prediction is concerned with determining the expected, valid flow through the program. Distinguishing between (input-specific) legitimate dynamic control flow and an illegitimate counterpart is critical to thwarting attempts to construct a malicious security apparatus. A key goal of this work is to build such an oracle that takes into account input relatedness and predicts the functions that correspond to the legitimate control flow. As described earlier, the predictor guides the BlankIt runtime to dynamically load only the predicted functions while other functions are kept blank. We now describe how we construct such an oracle based on decision trees. 
\subsection{Overall Framework}
%Figure \ref{figure:framework} shows an overview of the entire prediction framework.
\begin{enumerate}
\item First the static divergence analysis finds statically non-divergent functions for which no prediction is needed.
\item We then instrument the reverse dominance frontiers (RDFs) corresponding to every argument at every call site to construct a call context for the library call.
\item The instrumented application is then profiled, and the call context and argument values are logged along with the call chain.
\item This profile information is fed into a machine learning model that constructs a decision tree to predict the call chain depending on the call context. 
\item The generated decision tree is then instrumented into the application at the application call site for prediction.
\item Finally, constant folding and dead code elimination is carried out to remove redundant predicates. 
\item The instrumented application then feeds the predicted call chain into the BlankIt runtime system.
\end{enumerate}
% We now describe each of the phases in detail. 

\subsection{Static Call Flow Divergence Detection}
In the first step of the prediction framework, we find library functions that don't have statically divergent call flow. In other words, we identify functions in which the call sites do not
reside in a different control flow path. If a library function is non-divergent, then all
the dynamic calls of the function will result in the same call chain.
We find non-divergent functions by checking if every call site within a function post
dominates the entry block of the function as shown in Algorithm \ref{SDD}. If the callee is
not on a control divergent path but is itself divergent, then the caller is also marked as a divergent function, and all other functions that call this function are also marked as divergent functions.

\begin{algorithm}
\caption{Static Divergence Detection}\label{SDD}
\begin{algorithmic}[1]
\Procedure{non-divergent}{Function F}
 \State {$\text{$BB_{entry}$} \gets \text{entry basic block in $F$} $}
\BState \emph{for each Call site $C_{i}$}:
 \State {$\text{$f_{ci}$} \gets \text{function called in  $C_{i}$} $}
\If {$ \text{ non-divergent($f_{ci}$) \&\& $C_{i}$ postdominates $BB_{entry}$} $}
\State $\text{F is non-divergent}$
\EndIf
\BState \emph{End for}
\EndProcedure
\end{algorithmic}
\end{algorithm}

In glibc, we found that among all the functions analyzed, roughly 27\% of the functions are
statically non-divergent. Such functions exhibit a constant single call-graph that can be
statically determined for which no dynamic prediction is required; however, dynamic prediction
is a must for the remaining 73\% of call sites. 

\begin{table}[h!]
 \centering
 \begin{tabular}{|l|l|l|}
   \hline
   \textbf{Library} & \textbf{\#Divergent} & \textbf{\#Non-divergent}\\
   \hline
   \hline
   glibc-2.26 	&	1985		&	737  \\
   \hline
  \end{tabular}
  \caption{Static divergence in glibc}
  \label{table:SDD}
\end{table}

 \subsection{Reverse Dominance Frontier-Based Prediction}
 After generating a database of statically non-divergent functions, we now look at predicting
 the call graph for the diverging ones. Normally, whole program path profiling would need
 instrumentation of every branch. This would result in a path prefix of a fixed window $W$,
 entailing the context of the last $W$ branches. This path prefix can then be used for
 prediction of the call chain within the function call. We leverage the fact that library
 functions are usually control divergent due to their arguments and contend that the static
 context required to predict the call chain is the control dependence of the arguments passed
 to the library.
 
\begin{figure}[h]
       \centering
        \includegraphics[scale=0.1]{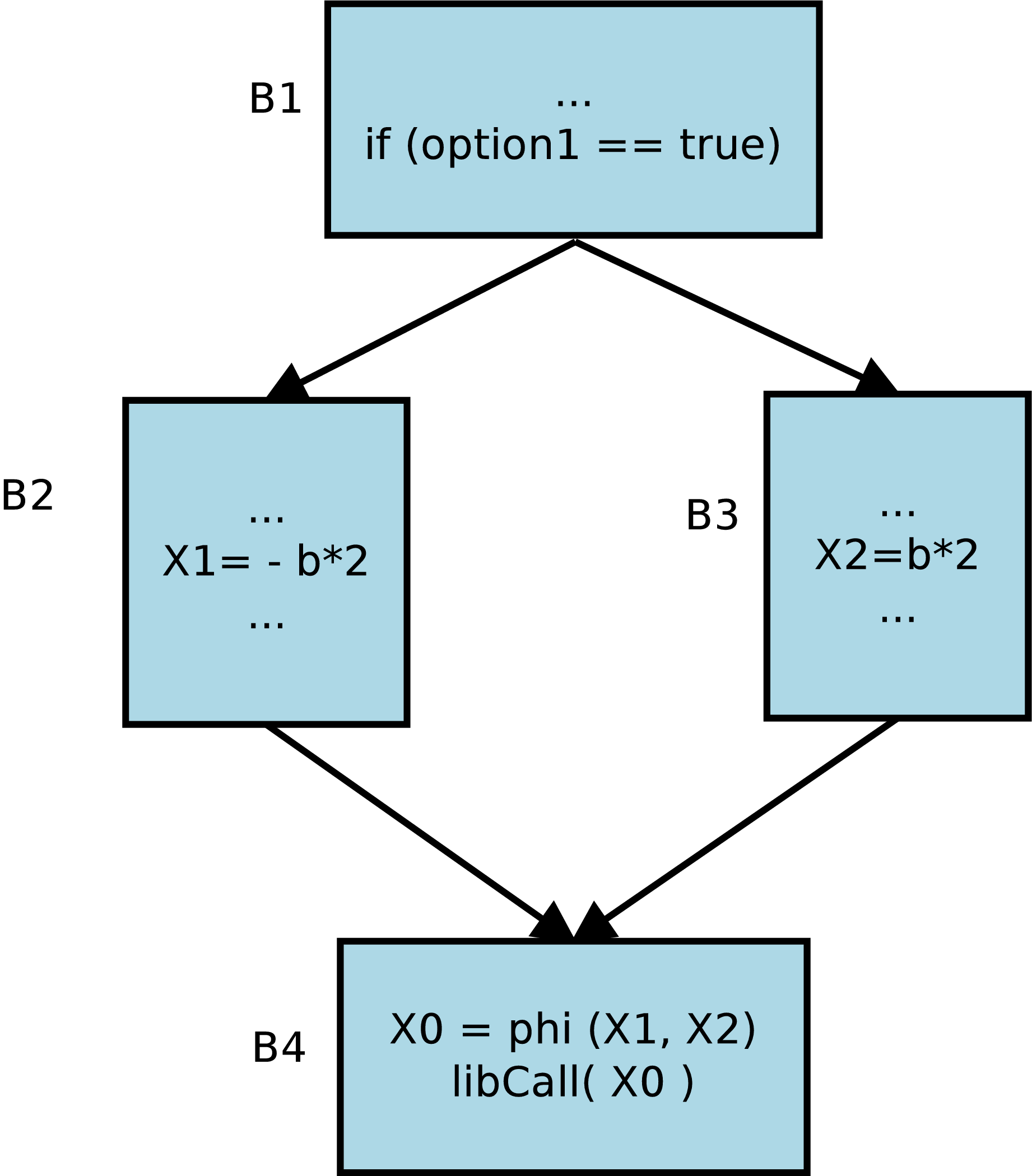}
        \caption{Simple control dependence example}
        \label{figure:rdf-1}
\squeezeup
\end{figure}

 This leads us to investigate the reverse dominance frontier (RDF) of the arguments of a call
 site, because the divergence in the library call path can be caused by the different
 definitions of arguments (determined by their respective RDF) reaching the call site. For
 example, in Figure \ref{figure:rdf-1}, a library call to $libCall()$ has one argument $X0$,
 which is defined as the phi instruction of arguments $X1$ and $X2$. If the basic block $B2$
 is executed, $X0$ has a negative value, but if the basic block $B3$ is executed, $X0$ has a
 positive value. In this case, both $B2$ and $B3$ are control dependent on $B1$. Thus the
 branch condition at $B1$ actually decides whether a positive or negative value is passed to
 $libCall()$. If the execution path and the call chain within the $libCall()$ function differs
 for positive and negative arguments, then we can learn to predict the call chain right after
 $B1$. The intuition here is that the call chain inside a library call will be dependent on arguments sent to the library, which in turn could be statically separated into corresponding
 SSA variables of a backward slice and finally guarding control dependent branches.

With RDF based prediction, the context leading up to a call site consists of only the last
executed RDF for every function argument. This minimal instrumentation has enough information
to predict the call chain within a function with reasonable accuracy. The only extra
instrumentation required is for correlating every branch with its corresponding function argument, and for each function argument to be instrumented with its RDF. When the profiling is done, the context for each call site is constructed separately, depending on the RDFs of its arguments. The instrumentation methodology is outlined in Algorithm \ref{rdf-instrument}.

\begin{algorithm}
  \caption{RDF-based Instrumentation }\label{rdf-instrument}
  \begin{algorithmic}[1]
    \Procedure{instrument\_function}{$Func$} %\Comment{The g.c.d. of a and b}
      
        \State $BBs\_to\_instrument\gets Null$   

      \ForAll{\texttt{ $callsite_i$ $\in$ $Func$}}
       \State {$\text{$F_{ci}$} \gets \text{function called in  $C_{i}$} $}
       \ForAll{\texttt{ argument $A_i$ $\in$ $F_{ci}$}}
       \State {$\text{$def\_worklist$} \gets \text{ get definition of $A_i$} $}

      \EndFor
      \EndFor
       \ForAll{\texttt{ instruction $def_i$ $\in$ $worklist$}}
       \State {$phi\_instr\_set \gets$ \Call{trace\_parent\_phi}{$def_i$} } \\  \Comment { call returns a set of phi instructions }
       
      \EndFor
       \ForAll{\texttt{$\phi_i$ $\in$ $phi\_instr\_set $}}
       \ForAll{\texttt{$incoming\_BB_i$ $\in$ $\phi_i$}}
        \State {$rdf\_set \gets$ \Call{get\_RDF}{$incoming\_BB_i$} }         
      \EndFor
      \EndFor
       \ForAll{\texttt{$rdf\_BB_i$ $\in$ $rdf\_set$}}
       \ForAll{\texttt{Successor $Succ_i$ $\in$ $rdf\_BB$}}
        \State {$BBs\_to\_instrument \gets$ $Succ_i$ }          
      \EndFor
        
      \EndFor
      \State \textbf{return} $BBs\_to\_instrument$
    \EndProcedure
    
    \Procedure{trace\_parent\_phi}{$instruction$} %\Comment{The g.c.d. of a and b}      
      \State $phi\_instr\_set \gets Null$   
      \ForAll{\texttt{ Operands $op_i$ $\in$ $instruction$}}
      \If {$op_i$ is\_a $phi$ instruction}
      \State {$phi\_instr\_set \gets op_i$ }
      \Else 
      \State {$phi\_instr\_set  \gets$ \Call{trace\_parent\_phi}{$op_i$} }  
      \EndIf            
      \EndFor
      \State \textbf{return} $phi\_instr\_set$
      \EndProcedure
  \end{algorithmic}
\end{algorithm}

\subsection{Argument Value-Based Prediction}
The control path in library functions can diverge because of the various values of the
arguments that are reaching the call sites in the application from the same RDF. Since the RDF
remains constant, RDF based prediction might not be sufficient. For example, we have observed
that math library functions like \texttt{sqrt(x)} and \texttt{exp(x)} have different call
chains depending on the range of the parameters. In order to handle such cases, we must
classify the values of the function arguments that reach from a single RDF, so we capture the
values of the arguments by profiling and appending the training model with the value. Once we
have the data on which we want to learn and predict the behavior of the library call graph at
each call site, we feed the data to a machine learning model.  
This value-based profiling also captures the values of the function pointers. Thus if a
function pointer is passed as an argument to another function, then we are able to predict the
function invocation based on the function pointer values.

\subsection{Implementation}
We used LLVM to insert instrumentation calls for recording the call context for every library
call made within the application. The instrumented application was run using Pin JIT to
generate the profile trace. The trace is then parsed to create a csv file with the training
data. 
We use the Python scikit \cite{scikit-learn} library to implement the machine learning model.
The decision tree API \\ \texttt{sklearn.tree.DecisionTreeClassifier}  is used to learn the
model and save it for later use. The python script reads in the tables produced from the
profiling phase. The learned decision tree model is then written to file as a special text
format. That final text file is read by an LLVM pass, which embeds the decision tree within
the application (i.e. inserts prediction calls for the runtime system).
We found experimentally that a max tree depth of 10 provided good accuracy with acceptable
peformance. Higher depth is only required if there are too many rows with the same features
but different outcomes, but adds the dilemma of overfitting the training data. We did not use boosting.
Please refer to the appendix section~\ref{sec:dtl} for additional details on decision tree learning.
\section{Evaluation}
\label{sec:evaluation}
We evaluate BlankIt using SPEC CPU 2006. SPEC has several properties that make it a good candidate for exercising and evaluating BlankIt. First, it is performance-oriented, so we can measure runtime overheads on realistic CPU-intensive benchmarks. Second, it has a standard library profile, which carries a relatively large CVE list. Lastly, call graphs of SPEC benchmarks and their supporting libraries are significantly dense and divergent, which are needed to stress the prediction framework. %(Table \ref{table:graph_metrics_table}).
As an example, SPEC benchmarks invoke 63 glibc unique functions at runtime on an average;
this leads to a 171-node, 378-edge static callgraph with a max callchain depth of 7
(and in this case a function set of which only 5\% are non-divergent) (see Table \ref{table:graph_metrics_table}).
A previous study~\cite{Gove:2007:ECT:1241601.1241624} demonstrates input similarity based on basic block and
branch execution frequencies, but the feature vector used for decision tree includes RDF and value separation
(with the intuition that the call chain inside the library is a function of the values of its arguments).
% In some cases, these arguments are influenced by the RDF in the program whereas in some others they are not.
While the RDF can have a relationship with the similarity measure based on the branch frequency in
~\cite{Gove:2007:ECT:1241601.1241624}, the values do not. Also, we predict the paths within the libraries, which
are not analyzed by this work.
Moreover, real-world applications have inputs that stress the program paths that can be used for training.

\begin{table}[h!]
 \centering
 \begin{tabular}{|l|l|l|l|l|}
   \hline
   \textbf{SPEC benchmark and glibc graph metrics} & \textbf{Avg}\\
   \hline
   \hline
    1. \# dynamically called glibc funcs from SPEC & 63 \\
    2. \# statically reachable funcs in glibc based on (1) & 171 \\
    3. \# callgraph edges based on static callgraph in (2) & 378 \\
    4. \# non-divergent functions in (2) & 8 \\
    5. max static callchain depth of (2) & 7 \\
   \hline
  \end{tabular}
  \caption{Key graph metrics for SPEC CPU 2006 and glibc}
  \label{table:graph_metrics_table}
   \squeezeup
%   \squeezeup
\end{table}

With the exception of the linker shared object and the vdso image, we have instrumented all the benchmarks' shared objects. Table \ref{table:shared_objects_table} shows the distribution of shared objects across the benchmarks. The most important is glibc-2.26\cite{glibc}, a fundamental library for programs written in C and compiled on GNU/Linux systems. The library provides POSIX, BSD, OS-specific and other APIs. Some of the basic facilities include print, login, and crypt. Thus, glibc supports a wide range of applications, which is borne out in the table (100\% dependence across all benchmarks). Historically, it has also been targeted and exploited heavily. The other libraries are libm 2.26, libgcc\_s from the libgcc1 package v1:7.2.0-1ubuntu1~16.04, and libstdc++ 6.0.24 which are also tackled by BlankIt. Our runtime uses Pin v3.6. We have built SPEC with version 5 of LLVM, which fails to compile 400.perlbench. It generates multiple definitions for \texttt{gnu\_dev\_major}, \texttt{minor}, and other functions, so we have elided treatment of it here. All other C/C++ benchmarks are presented. The raw decision trees have a lot of compile time constant checks which after constant folding reduces the tree to few \textit{if else} checks thus increasing binary size only modestly.  All runtime overhead experiments are on an Ubuntu 16.04.3 LTS machine with an AMD Ryzen 7 1800X 8-Core 3.6 GHz processor and 32 GB DD4 2666 MHz RAM. All the SPEC benchmarks were trained on ``test'' (small) and ``train'' (medium) input data sets and were regression tested using the (large) ``ref'' input data sets. All runtime experiments are averaged over three runs.

\begin{table}[h!]
 \centering
 \begin{tabular}{|l|l|l|l|l|}
   \hline
   \textbf{Benchmark} & \textbf{libc} & \textbf{libm} & \textbf{libgcc} & \textbf{libstdc++}\\
   \hline
   \hline
    401.bzip2	 &	\checkmark &	&	 &	\\
    403.gcc	 &	\checkmark & &	 &	\\\
    429.mcf	 &	\checkmark &	 &	 &	\\
    433.milc	 &	\checkmark &	\checkmark &	 &	\\
    444.namd	 &	\checkmark &	\checkmark &	\checkmark &	\checkmark\\
    445.gobmk	 &	\checkmark &	\checkmark &	 &	\\
    450.soplex	 &	\checkmark &	\checkmark &	\checkmark &	\checkmark\\
    453.povray	 &	\checkmark &	\checkmark &	\checkmark &	\checkmark\\
    456.hmmer	 &	\checkmark &	\checkmark &	 &	\\
    458.sjeng	 &	\checkmark &	 &	 &	\\
    462.libquantum	 &	\checkmark &	\checkmark &	\checkmark &	\\
    464.h264ref	 &	\checkmark &	\checkmark &	 &	\\
    470.lbm	 &	\checkmark &	\checkmark &	 &	\\
    471.omnetpp	 &	\checkmark &	\checkmark &	\checkmark &	\checkmark \\
    473.astar	 &	\checkmark &	\checkmark &	\checkmark &	\checkmark\\
    482.sphinx3	 &	\checkmark  &	\checkmark &	 &	\\
    483.xalancbmk	 &	\checkmark &	\checkmark &	\checkmark &	\checkmark\\
   \hline
  \end{tabular}
  \caption{Shared objects in SPEC CPU 2006 benchmarks}
  \label{table:shared_objects_table}
   \squeezeup
%   \squeezeup
\end{table}

\subsection{Security}
What constitutes an attack surface, especially when preparing for unknown future attacks, is difficult to capture. We present three results for reducing dynamic linked functions. Taken together, they offer three different axes for understanding how BlankIt is behaving. They summarize (1) the number of functions that are exposed, (2) the number of gadgets in a known type of attack that are exposed, and (3) the number of known functions from a vulnerable list that are exposed.

The first measure is a dynamic metric that describes the maximum number of library functions exposed at any given time during execution. This can be described by the following formula:
\begin{equation}
exposed = p + s + c
\end{equation}
where \textit{exposed} is the maximum number of loaded functions at runtime, \textit{p} is the number of functions that Pin is unable to instrument and thus must be left as they are (without blanking them), \textit{s} is the number of functions that are less than 14 bytes and therefore too small for blanking, and \textit{c} is the maximum-length call chain for a given benchmark that our framework dynamically loads at any callsite during the execution. In other words, the number of exposed functions in the worst case is the maximum number of unblanked functions that could be leveraged in some attack. The percent reduction of code surface is then given by:
\begin{equation}
% \squeezeup
reduction = \frac{\sum_{l} n_l - exposed}{\sum_{l} n_l} * 100
% \squeezeup
\end{equation}
where \textit{l} is some library and \textit{n\textsubscript{l}} is the total number of functions available at runtime in some library \textit{l}.
This metric does not capture anything about the inherent weaknesses or strengths of the exposed functions - it only captures that they are exposed. Table \ref{table:attack_surface} presents this result in the first column as ``\% reduction in dynamic linked functions''. This worst-case reduction is, on average, 97.1\%.

\newcommand{\breakablecell}[2][c]{%
  \begin{tabular}[#1]{@{}c@{}}#2\end{tabular}}
\begin{table}[h!]
 \centering
 \begin{tabular}{|l|l|l|l|}
   \hline
   %\textbf{Benchmark} & \textbf{\% Reduction in DLF} & \textbf{\% ROP Gadget Reduction}\\
   \textbf{Benchmark} & \breakablecell[t]{\bf \% \bf Exposed \\ \bf Code\\ \bf Surface\\ \bf Reduction} & \breakablecell[t]{\bf \% \bf ROP \\ \bf Gadget\\ \bf Reduction} & \breakablecell[t]{\bf \% \bf glibc \bf CVE \\ \bf Function\\ \bf Reduction}\\
   \hline
   \hline
    401.bzip2	 &	97.7 & 98.9 & 95.7\\
    403.gcc	 &	97.2 & 97.8  & 95.7\\
    429.mcf	 &	94.5 & 95.9  & 95.7\\
    433.milc	 &	98 & 98.5  & 95.7\\
    444.namd	 &	97 & 96.4  & 93.6\\
    445.gobmk	 &	95.7 & 96.2  & 93.6\\
    450.soplex	 &	97.4 & 99.3  & 97.9\\
    453.povray	 &	96.9 & 97.7  & 93.6\\
    456.hmmer	 &	97.9 & 97.9  & 93.6\\
    458.sjeng	 &	97.8 & 98.9  & 95.7\\
    462.libquantum	 &	97.9 & 98.6  & 95.7\\
    464.h264ref	 &	97.9 & 98.7  & 95.7\\
    470.lbm	 &	97.8 & 98.6  & 95.7\\
    471.omnetpp	 &	95.6 & 96.9  & 95.7\\
    473.astar	 &	96.6 & 96.6  & 95.7\\
    482.sphinx3	 &	97.6 & 98.2  & 95.7\\
    483.xalancbmk	 &	96.9 & 98.1  & 91.5\\
   \hline
  \end{tabular}
  \caption{Reductions metrics (DLF = Dynamic Linked Function)}
  \label{table:attack_surface}
\squeezeup
\end{table}
The results show a very high percentage reduction; significantly more than any static or dynamic technique can achieve.

Return-oriented programming depends on ROP gadgets in the code in order to carry out an effective attack. To measure the reduction in gadgets, we leveraged ROPgadget \cite{ropgadget}, an analysis tool for enumerating the ROP gadgets in a binary. Because BlankIt is a dynamic technique, the number of gadgets varies over time. Thus, similar to before, we choose the worst-case scenario at any given point in the program and report its ROP gadget reduction. As before, we calculate the maximum number of exposed gadgets (similar to equation 1), and then calculate the reduction by summing over the number of gadgets in the text section in each library, subtracting out the exposed gadgets, and gathering the percentage (as in equation 2).
%\begin{equation}
%exposedg = pg + sg + cg
%\end{equation}
%where the \textit{exposedg} is the maximum number of exposed gadgets at runtime, \textit{pg} is the number of gadgets in functions that Pin is unable to instrument, \textit{sg} is the number of gadgets in functions that are less than 14 bytes and therefore too small for blanking, and \textit{cg} is the number of gadgets in the maximum-length call chain for a given benchmark. The reduction calculation follows the same 
Table \ref{table:attack_surface} shows the benefits of BlankIt on ROP gadget reduction. The average is 97.8\%.
Please refer to the appendix section ~\ref{sec:gq} for details on the quality of these gadgets.
%We also looked at gadget ``quality'' \cite{gadgetquality}, but it is difficult
%in the case of BlankIt's dynamic runtime to make a round statement about it. The longest call chain
%may have the least gadget reduction but not necessarily the worst gadget quality. Identifying
%the call chain with the worst gadget quality does not necessarily mean, however, that that call
%chain is unsafe, because it may also be so heavily stripped of ROP gadgets that there is no risk.

The last metric is a measure of the functions in glibc CVEs that are removed by BlankIt. We reviewed the list of all 95 CVEs for glibc, which reach back to year 2000, and we identified all unique functions mentioned in the descriptions. Of these, we identified 47 that are unequivocally loaded by glibc in the SPEC suite. That is, there must be an exact match in the    
dynamically loaded list of glibc functions for a function related in a glibc CVE to be considered. For example, ``alloca'' is mentioned in CVE-2015-1473, but it is not explicitly exported with that name and so 
is discarded, even though there are multiple allocation functions. Then the number of exposed CVE functions is obtained as follows:
\begin{equation}
exposedcve = p + s + a
\end{equation}
where \textit{p} and \textit{s} are uninstrumented due to Pin or too small to instrument (as before), but \textit{a} represents \textit{any} function that is called that is in the CVE function list. The percentage is then taken out of the 47 CVE functions. This metric should not be misunderstood as the number of CVEs since 2000 that are thwarted by BlankIt. Rather, it classifies functions as vulnerable based on their CVE history, and then asks how many such functions are exposed under BlankIt running a common benchmark. The results are shown in the last column of Table \ref{table:attack_surface}. The average reduction is 95.1\%.

%\subsection{Prediction Accuracy}
%Prediction accuracy is hand-in-hand with security, but we separate it for clarity. 

\subsection{Runtime}
\begin{figure*}[h]
    \centering
    \includegraphics[width=15cm,height=7cm]{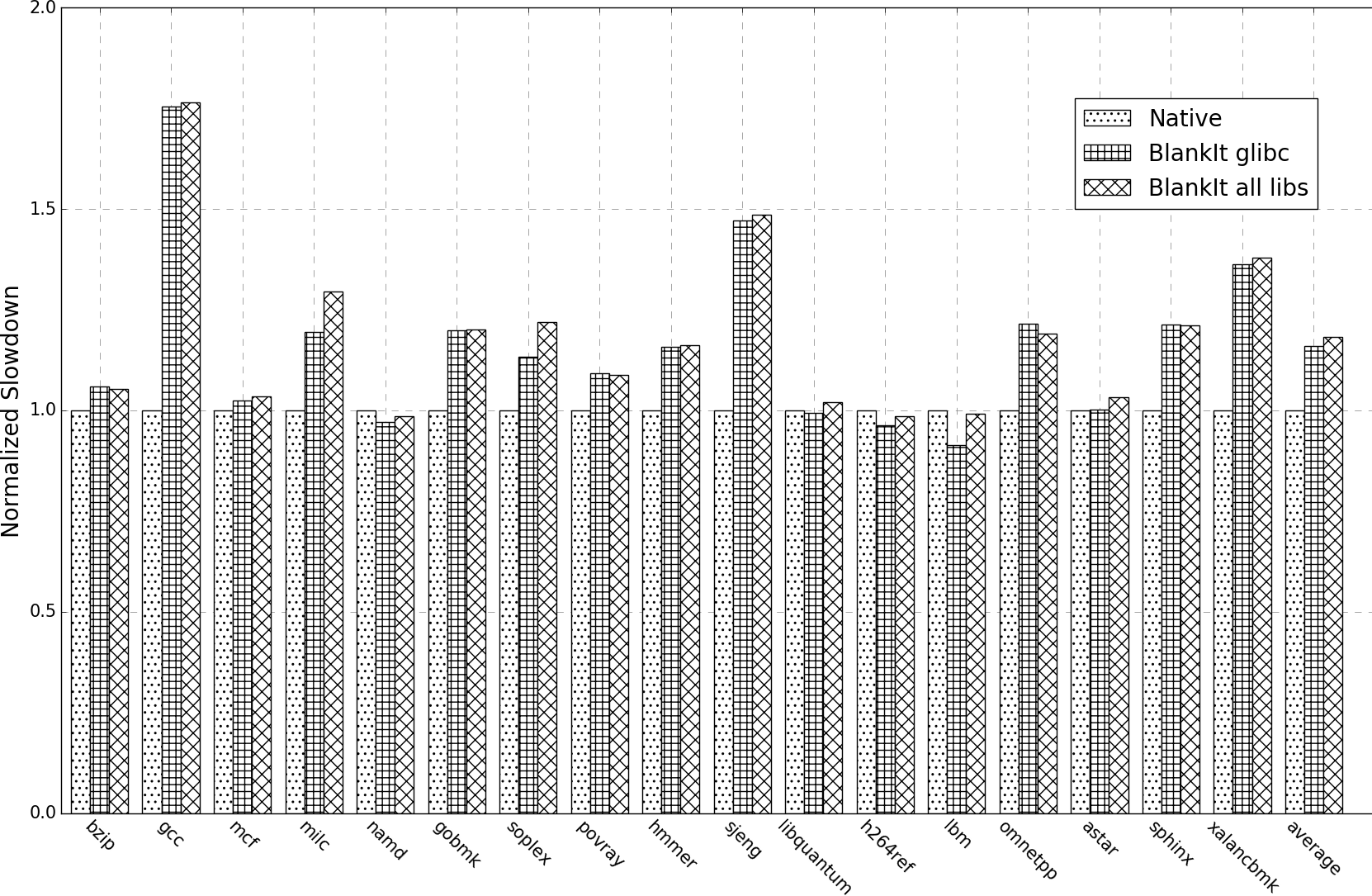}
    \caption{Runtime slowdown for BlankIt on SPEC 2006 CPU, normalized against native.}
    \label{figure:blankit_runtime_overhead}
\squeezeup
\end{figure*}

Figure \ref{figure:blankit_runtime_overhead} shows the runtime overhead. There are three bars per benchmark: native, BlankIt for glibc, and BlankIt for all libs. The most crucial library for security purposes is glibc (the middle bars for each benchmark in the graph). The average slowdown for a BlankIt-enabled application on only glibc is 16\%. There is a slight speedup in the case of lbm, which can be accounted for by variance. The rest perform reasonably well, with gcc being the worst at 1.76x. The slowdown for gcc (and similarly, for sjeng) is due to heavy usage of libc.  Runtime overhead is primarily due to library call frequency and not due to misprediction. Thus, while prediction accuracy was poorest for libquantum (discussed in the next section), the performance was good. In this BlankIt prototype the audit thread runs in parallel and ``keeps up'' with the mispredict frequency and does not add extra blocking. There are still open optimization opportunities to improve the performance, which may include compiler-based hoisting operations to pull prediction probes outside of loops.

The rightmost bars represent a BlankIt-enabled application on all libraries. In some cases, (bzip2, gcc, mcf, sjeng) the results are almost identical to a glibc-only result because they link against no other libraries. In other cases, adding the libraries costs little in overhead for the extra security. The average slowdown for BlankIt over all libs is 18\%, demonstrating that BlankIt likely scales well across programs' dynamic library sets.

\subsection{Prediction}

Our treatment of prediction includes our observations while working with the framework, accuracy results and analysis on SPEC, and then a breakdown of the kinds of mispredictions that were measured.

% \squeezeup
\subsubsection{Observations}
In the case of library functions, especially in languages like C that do not have a context for library function calls, we have observed that the set of functions 
called within the library can be predicted based on the dynamic calling context. The necessary dynamic calling context is discovered by the decision tree which can  
as simple as just the name of the function and/or the call site. 
%For example, functions like \texttt{malloc}, \texttt{calloc}, and \texttt{strlen}, which are some of the most frequently called library functions, can be predicted 
%with high accuracy just based on the call site and function name. In other words, irrespective of the arguments passed, the function almost always has a fixed set of internal calls.

%In certain other cases, the call site must be used by the decision tree to predict the call chain with high accuracy. To rephrase, calling the same function at two different sites sometimes results in two different call chains. These cases are more complex than function name alone, but they are still predictable.
%Concretely, \texttt{printf} may call a different set of functions depending on its call site, but the call chain is fixed per call site. In this case, the decision tree can map the call site to the call chain, without using the function name, since the function name is fixed once the call site is determined. 
However, some times the call chain is dependent on the arguments passed to the function, which could include a function pointer. For example, there might be a math library function that calls a different set of internal functions based on the arguments. The decision tree is able to handle such cases by training on the values passed to the function. It is able to model the different call chains that are invoked based on the range of values of the function parameters encountered during profiling. This value-based approach handles function pointers as any other value. That is, if a function takes a function pointer as an argument, then based on the value of the pointer, different call chains will be invoked. 
If the profiling data has all possible values that a function pointer can take, then our decision tree is able to train on them to predict the right call chain, depending on the function pointer value. 

The decision tree is currently unable to capture other use cases like the presence of multiple threads or certain special system states or error conditions which were not triggered during profiling.

% \squeezeup
\subsubsection{Accuracy and Audit overhead}

Table \ref{table:prediction-accuracy} shows the prediction accuracy and audit overhead for the SPEC2006 benchmark suite.
%All the SPEC benchmarks were trained on ``test'' (small) and ``train'' (medium) input data sets and were regression tested using the ``ref'' (large) input data sets.
Over half of the accuracies are 97\% or greater; three are 100\%; and the average is 94.3\%. As a matter of fact, the decision tree's prediction ability was tested for predicting every function in SPEC2006 at every call site and was found to be very close to the one reported here for the libraries.

\begin{table}[h!]
 \centering
 \begin{tabular}{|l|l|l|}
   \hline
   \textbf{Benchmark} & \bf \bf \% Accuracy &  \bf Audit ($\mu s$) \\
   \hline
   \hline
    401.bzip2	 &	91 & 287  \\
    403.gcc	 &	99 & 12 \\
    429.mcf	 &	94 & 92 \\
    433.milc	 &	100 & 27 \\
    444.namd	 &	99 & 11 \\
    445.gobmk	 &	84 & 8 \\
    450.soplex	 &	92 & 26 \\
    453.povray	 &	97 & 6 \\
    456.hmmer	 &	98 & 6 \\
    458.sjeng	 &	97 & 27 \\
    462.libquantum	 &	60 & 41 \\
    464.h264ref	 &	100 & 28 \\
    470.lbm	 &	98 & 171 \\
    471.omnetpp	 &	99 & 6 \\
    473.astar	 &	100 & 58 \\
    482.sphinx3	 &	99 & 11 \\
    483.xalancbmk	 &	96.9 & 18 \\
   \hline
  \end{tabular}
  \caption{Call Chain Prediction Accuracy and Audit Overhead}
  \label{table:prediction-accuracy}
  \squeezeup
%   \squeezeup
\end{table}

One of the important reasons for very high prediction accuracy stems from the fact
that libraries follow software design patterns; the use cases of libraries are therefore
finite and can be siloed. Our decision trees are able to learn these patterns
which are input-related and during regression testing are able to pinpoint
the silo and its underlying set of function calls. 
The only application for which the decision tree accuracy was relatively
low is libquantum. For this application, there was a call chain involving cfree
and cfree followed by munmap in train and ref inputs, respectively, which is one
of the major sources of misprediction.
The reasons for these differences are due to the size of memory that is being
freed which is not captured by the input oriented mechanism based on RDF
and argument value separation.

The audit overhead is reported as the geomean of mispredicted function runtime
within valgrind.  The results add confidence that running an audit in parallel
is tenable. The total library function call counts are not reported, but as a
typical example, lbm has 2,626,463 runtime library calls (and 98\% prediction accuracy).
lbm takes roughly 250s to execute under BlankIt. Thus, a 171$\mu s$ overhead per mispredict
falls well within the \~210 mispredictions that occur per second.
This shows that doing full program stops until Valgrind reports back is plausible.
% In fact, this could open the avenue for doing full program stops until valgrind reports back.
% For example, the library function \texttt{cfree} calls \texttt{munmap} during test runs but not during training runs. \texttt{munmap} is a function called to free memory-mapped addresses. Thus, our decision tree analysis cannot handle cases that were not encountered during the profiling and training.  

% \squeezeup
\subsubsection{Misprediction Breakdown}

%Table \ref{table:misprediction-classification} is a summary breakdown of misprediction into overpredictions (where more functions are loaded than were needed) and underpredictions (where more functions had to be loaded after a prediction was made). Overprediction could potentially be more problematic for security, as it increases the attack surface beyond what was needed at runtime. Note that in BlankIt, however, even for the cases where it occurs (gcc and soplex from this list), it is \textit{bounded}. The decision tree is computed pre-runtime, and all prediction sets (and lengths, and their attack surface exposure), are known before the application executes. And because the sets reside in a protected environment and invalid IDs cannot be passed from the application into BlankIt runtime, the overprediction security maintains runtime safety guarantees.

We further classify the mispredictions into overpredictions (where more functions were loaded than necessary) and underprediction (where lesser number of functions loaded than necessary). In almost all the cases, there were under-predictions (100\%) except in case of gcc (where the under and over predictions were almost equal) and soplex (where there were under-predictions 93\%). 
 The main reason is that when the decision tree encounters a call-site which was not exercised during training, it is extremely conservative and bases the prediction on whether the entry function was seen during training; due to these reasons it chooses a subset of functions that were callees of the entry functions. This in turn provides good security since no spurious functions are brought in that could increase the exposed code surface.
 
\subsection{Multithreading and Windows}
 BlankIt is not limited to single-threaded programs, neither in its general approach nor current implementation, nor is it OS-specific. BlankIt can leverage Intel Pin's threading support and add a synchronized blanking/un-blanking mechanism for parallel programs. The thread that blanks (or fills) a piece of code would have to acquire a lock for that particular code section to do so (page granularity in Linux). Similarly, BlankIt can also be ported to Windows without any design or conceptual changes: mprotect can be replaced by VirtualProtect, msvcrt.dll protected in place of glibc, etc.
    
% \squeezeup

% \begin{table}[h!]
%  \centering
%  \begin{tabular}{|l|l|l|p{12mm}|}
%   \hline
%   \textbf{Benchmark} &  \textbf{Over Predict} & \textbf{Under Predict} & \textbf{Total Mispredict} \\
%   \hline
%   \hline
%   % 401.bzip2	 &	91 \\
%     403.gcc	 &	9 & 7 & 16 \\
%     429.mcf	 &	0 & 854 & 854 \\
%     444.namd	 &	0 & 2199 & 2199  \\
%     445.gobmk	 &	0 & 3 & 3 \\
%     450.soplex & 20 & 308 & 328 \\
%     462.libquantum	 &	0 & 78 & 78 \\
    %464.h264ref	 &	100 \\
    %470.lbm	 &	98 \\
    %471.omnetpp	 &	99 \\
    %473.astar	 &	100 \\
    %482.sphinx3	 &	99 \\
    %483.xalancbmk	 &	96.9 \\
%   \hline
%   \end{tabular}
%   \caption{Prediction classification}
%   \label{table:misprediction-classification}
% \end{table}

%% THE above table has to go... just before the Related Work section
\section{Case Study}\label{case_study}
We study how a recent vulnerability in glibc, CVE-2018-11236~\cite{CVE-2018-11236} on buffer overflow
is handled by BlankIt. CVE-2018-11236 is a vulnerability that exists in glibc version 2.27 and earlier in
 the realpath function in stdlib/canonicalize.c. The function realpath intakes a pathname as one of its
 arguments. If a pathname that has length close to SSIZE\_MAX is passed as an argument, it causes a buffer overflow
 that overwrites the stack on line 191 in Code\ref{code:CVE}. This overflow can be exploited to
 jump to any ROP/JOP gadget and carry out a Turing-complete attack. With BlankIt, however, jumping
 to any arbitrary location is not possible, as the code sections
 are blanked. CFI mechanisms would limit arbitrary jumps, as well. However, in CFI the attacker
 can jump to a legit 
 function with exploitive inputs and carry out the attack. BlankIt catches the attack if the 
 call to this function is not predicted in the path. Further, if the function 
 realpath is never used, then BlankIt never loads it, catching any attempt to
 attack through the function.

\renewcommand{\lstlistingname}{Code}
\begin{lstlisting}[firstnumber=183,xleftmargin=.1\textwidth,caption={CVE-2018-11236},label=code:CVE]
 extra_buf = __alloca (path_max);
 len = strlen (end);
 if (path_max - n <= len)
 {
    __set_errno (ENAMETOOLONG);
    goto error;
 }
 /* Careful here, end may be a pointer into extra_buf... */
 memmove (&extra_buf[n], end, len + 1);
\end{lstlisting}
\squeezeup
% \squeezeup

\section{Related Work}

The two main lines of research that are closely related to security-focused debloating are software debloating and security vulnerability techniques.

%\subsection{Code Debloating}
 The embedded software community performed debloating based on link-time optimizations in order to
 reduce code size. As mentioned earlier, compiler based link-time optimizations involve statically
 determining a set of functions to be removed from the linked library. The key goal of such link time
 optimizations is to compact the code size ~\cite{Beszedes:2003:SCR:937503.937504, Debray:2000:CTC:349214.349233}. While such solutions~\cite{Muth:2001:ALO:370365.370382, Franz:1997:SB:265563.265576}  have shown to substantially reduce the code size, they are not geared
 towards solving the security issues due to  bloated libraries. This is because {\it static}  link
 time optimizations err on the conservative side. A load time mechanism such as piece-wise compilation~\cite{217642} reduces approximation but relies on a CFI mechanism for defense against the ROP gadget exploitation in the loaded functions.
%  They determine a superset, which means they potentially leave some functions in the library which could be dynamically unreachable during {\it any} execution of that application; such dynamically dead functions could contain vulnerabilities and provide entry points into the main application through the exploits of such vulnerabilities. Determination of such dead functions is an undecidable problem. Even if one manages to somehow solve the problem of deadness, a second major limitation of the above approaches still persists. These approaches find the sum total of every reachable functions through the above techniques at all call sites and leave those intact in the linked version of the library; thus, an ROP gadget can be constructed using the library calls across two call sites. 
%  These two limitations should prompt new directions rather than extensions of current practices. 
A new direction of code debloating based on programmer specifications  was developed in~\cite{Heo:2018:EPD:3243734.3243838}. These problems are different from the goal of BlankIt, which reduces the number of {\it dynamic functions} linked so that the possibilities of
constructing malicious programs diminishes significantly. 

 %% COMMENT: To save space, you may want to cut down the following summarizing as software engg efforts focused %% on program
 %% bloat but has nothing to do with security 
Debloating is an active research topic within the software engineering community and oftentimes with respect to performance. Mitchell et al. \cite{causes_of_bloat} investigated causes of bloat and how ``health signatures'' can describe memory footprints in a way that lends itself to value judgments about the strength of a software design. In a similar work, a tool called Yeti was developed to help identify costly data structures \cite{making_sense_debloat}. Regarding performance, Xu et al. \cite{analysis_debloat} argue that memory bloat is a more severe problem given reduced scaling of chip real estate due to Moore's law. Other research is on understanding excessive abstractions or object creation in Java and how these affect performance \cite{four_trends_debloat}; on identifying heavy computations that have little benefit (e.g. constructing a large object only to check its size) \cite{finding_utility_debloat}; on debloating container objects due to their negative impact on performance \cite{container_debloat}; and on reusing data structures rather than recreating them to boost performance \cite{finding_reusable_debloat}. While this work has paved some of the way for debloating, it is not geared towards security and thus can not be directly applied to the problem of reducing exposed code surface.
%% COMMENT: What about memory safety work? Nagarakatte et al Huge overheads - we invoke audit on demand - cutting down unnecessary overhead

%\subsection{Security Vulnerabilities}
The lack of memory safety in  languages such as C/C++ has been a pernicious and long-lasting problem,
and a vast number of potential solutions have been proposed. Real world exploits show that all
currently deployed protections can be defeated, and in fact new vulnerabilities are reported
frequently ~\cite{stackBufferOverflow, DNSBufferOverflow, CVE}.
% While the attacks start by exploiting
% a weak memory model, they exploit function pointers or more generally control pointers, and spread by
% hijacking the control flow, leading it where it is not intended to go.
Memory safety mechanisms can safeguard against many of the memory corruption attacks, however, is $\geq 2X$ slower in case of source based memory safety techniques like CETS~\cite{Nagarakatte:2010:CCE:1806651.1806657} and $\geq 10X$ slower in case of binary based ones like Dr.Memory~\cite{Bruening:2011:PMC:2190025.2190067}.
There has been work on control pointer integrity (CPI) checking 
~\cite{Kuznetsov:2014:CI:2685048.2685061,Bittau:2014:HB:2650286.2650800}, Address Space Layout
Randomization ~\cite{Snow:2013:JCR:2497621.2498135}, as well as control flow integrity (CFI) checking
~\cite{cfi,control_flow_bending,bincfi,DBLP:conf/asplos/GeCJ17,7920853}.
% Ge et al. (ASPLOS 2017) proposed Griffin, and Liu et al. proposed FlowGuard (HPCA 2017), which both leverage Intel Process Trace hardware for fine-grained CFI. These techniques still suffer from CFI attacks mentioned previously, such as Control Jujutsu and control flow bending. BlankIt is capable of catching such attacks when they fall within a blanked region.
In spite of such advancements, in ~\cite{control-jujutsu} authors show an attack called Control Jujutsu that exploits the imprecision of scalable pointer analysis to bypass fine-grained enforcement of CFI.
% The attack uses Argument Corruptible Indirect Call Site (ACICS) that can hijack control flow to achieve remote code execution while still respecting control flow graphs generated using context- and field-sensitive pointer analyses. It is shown that preventing Control Jujutsu by using more precise pointer analysis algorithms is difficult for real-world applications; it is also difficult to construct the precise CFGs for several reasons. Furthermore, as pointed out in the introduction,
Control-Flow Bending~\cite{control_flow_bending} (CFB) showed that even a fully-precise
static CFI cannot defend against many non-control data attacks.
% CFB demonstrated that with
% the availability of a dispatcher function, commonly available in glibc, a CFI mechanism can
% be fooled to successfully carry out an attack.
In addition to these attacks, there are many DoS (denial of service) and privilege escalation attacks documented in ~\cite{CVE} that are unrelated to the issue of control flow and could have severe implications for the security of the linked application.As discussed earlier in the introduction, BlankIT prevents some attacks that escape detection techniques based on CFI and CPI.

\section{Conclusion}

In this work, we propose and show how to effectively implement a scheme
of demand-driven loading to reduce the exposed code surface of
vulnerable linked libraries. We achieve this by capturing control dependence of library call sites (via reverse dominance frontiers) and by training decision trees to establish input-relatedness. We also devise a highly effective, very low overhead, binary-level mechanism. It dynamically loads the needed functions at a given call site and blanks out the unneeded ones, which significantly weakens an attacker's ability to construct gadgets.
Our results on the full SPEC2006 benchmark suite (evaluated with its 4 major user-level libraries, libc,
libm, libgcc and libstdc++) are very strong: the prediction accuracy in most cases being at least 97\%,
and dynamic linked functions and ROP gadget reduction likewise being at least 97\%. On average, the runtime overhead with BlankIt  is 18\%, across full set of SPEC benchmarks which is within tolerable limits.
BlankIt also catches few cases of non-control data attacks and some which are not tackled by CFI and CPI in general. BlankIt has all these benefits without
recompiling any linked binaries.
% There is no need for recompilation of any linked binaries.
% We conclude that the BlankIt system provides an extremely high amount of attack surface reduction at small to moderate performance penalties. % (barring the outliers). 
%with an extremely low (typically within 3\%) misprediction rate; thus, we believe our scheme is practicable. Our future work on this topic involves implementing optimizations for decision trees and hoisting both loading and blanking calls at appropriate program points to reduce the runtime overheads for the two outliers.

% \section{Future Work}
% The approximation in decision trees towards call graph prediction is a limitation of decision trees which can 
% predict path with ease as shown in this paper. In our future work, we plan to study these limitations, classify and address them.

%-------------------------------------------------------------------------------
%\bibliographystyle{plain}
%\bibliography{texes/references}

\section{Appendix}\label{sec:Appendix}
\subsection{Quality of Gadgets}\label{sec:gq}
\begin{table}[h!]
 \centering
 \begin{tabular}{|l|l|l|}
   \hline
   \textbf{Benchmark} & \textbf{Gadget quality} & \textbf{BlankIt Gadget quality} \\
   \hline
   \hline
    401.bzip2	 &	0.81 & 1.25		\\
    403.gcc	 &	0.81 &	1.07	\\\
    429.mcf	 &	0.81 &	1.14		\\
    433.milc	 &	0.75 &	1.29		\\
    444.namd	 & 	0.77 &	0.97		\\
    445.gobmk	 &	0.75 &	0.84		\\
    450.soplex	 & 0.77	 &	0.96		\\
    453.povray	 & 0.77	 &	1.12		\\
    456.hmmer	 &	0.75 &	1.19		\\
    458.sjeng	 &	0.81 &	1.15		\\
    462.libquantum	 & 0.73	 &	 0.79		\\
    464.h264ref	 &	0.75 & 1.15			\\
    470.lbm	 &	0.75 & 1.35		\\
    471.omnetpp	 &	0.77 &	1.08		 \\
    473.astar	 & 0.77	 & 1.15			\\
    482.sphinx3	 &	 0.75 &	0.97		\\
    483.xalancbmk	 &	0.77 &	0.98		\\
   \hline
  \end{tabular}
  \caption{ROP Gadget Quality in SPEC CPU 2006 with and without BlankIt (higher indicates better security)}
  \label{table:gality-table}
   %\squeezeup
  %\squeezeup
\end{table}

%libc = 0.81
%libc, libm = 0.75
%libc, libm, libgcc = 0.73
%libc, libm, libgcc, libstdcpp = 0.77

%401.bzip2 1.25
%403.gcc 1.07
%429.mcf 1.14
%433.milc 1.29
%444.namd 0.97
%445.gobmk 0.84
%450.soplex 0.96
%453.povray 1.12
%456.hmmer 1.19
%458.sjeng 1.15
%462.libquantum 0.79
%464.h264ref 1.15
%470.lbm 1.35
%471.omnetpp 1.08
%473.astar 1.15
%482.sphinx3 0.97
%483.xalancbmk 0.98

% girish: the gality reference in the bib is "gadgetquality" -- got it
We would like to present details on the quality of gadgets with respect to BlankIt.
Gality~\cite{gadgetquality} is a tool that measures the quality of the gadgets for the ease of constructing an
attack and provides an overall score. In details, the tool first categorizes the gadgets,
finding the distribution of different gadgets available in the arsenal of the attacker.
The tool also checks the side effects of a gadget. A side-effect is 
measured by the number of preconditions that must be satisfied for using a gadget. In 
other words, a high quality gadget is one with no pre-conditions or that does not 
affect any other register or memory by overwriting the value and thus giving away 
the attack. The Gality score starts at 0 and is increased for side-effects and 
pre-conditions. Thus higher score indicates worse gadget quality hence better security.
As shown in Table \ref{table:gality-table} BlankIt consistently has higher scores thus 
signifying that BlankIt offers better security.

\subsection{Non-Control Data Attack Defense}\label{sec:nc}
We demonstrate non-control data attack that is mitigated by BlankIt.
In Figure \ref{figure:misprediction_phase}, the
arguments of the library function are copied at the latest available point, which is at
either the PHI function
or the definition (whichever is the reaching definition). The copies are passed to the auditor
process during misprediction. Any input-based overflow within the library is caught by Valgrind.
However, if the input to the library is modified within the application and results in
a valid program flow, then Valgrind does not report an error.
Thus, in cases where the input to the library
is modified after
the latest availability point of an argument, BlankIt can catch the attack by
comparing the predicted call
chain with the call graph reported by the auditor process.  In the example, input 'a' is 
modified through buffer overflow in 'strcpy(b,input)' and results in a misprediction. However, the
predicted chain matches with the call graph reported by the auditor, thus catching the attack
and raising an alarm.
During actual misprediction, the Valgrind call chain matches the observed call chain and not the
predicted call chain and thus, declares it as legal execution.

\subsection{Decision Tree Learning}\label{sec:dtl}
The decision tree is a machine learning model based on inductive inference and which satisfies
BlankIt's requirement for quick runtime lookup. It is trained on a set of input attributes and
the observed output by correlating the observed output with a disjunction of conjunctions of
predicates on the input attributes \cite{Mitchell:1997:ML:541177}. 
The decision tree tests a predicate on one of the input attributes at every node. Depending on
the outcome of the predicate, one of the outgoing edges is followed to the next node and halts
when a leaf node is reached. A leaf node represents a class label, which is the prediction for
the given input.
In our model, the input is the context leading up to a call site, and the output is the call
graph of the callee function. 
The context at a call site is the call site itself, the function that is called, and the
function argument values. Given the context at a call site for a callee function $F$, the
model will predict the set of functions within the library called through $F$.
%%%%%%%%%%%%%%%%%%%%%%%%%%%%%%%%%%%%%%%%%%%%%%%%%%%%%%%%%%%%%%%%%%%%%%%%%%%%%%%%
\end{document}